\documentclass[conference]{IEEEtran-v17}

\IEEEoverridecommandlockouts
\usepackage{amssymb}
\usepackage{amsmath}
\usepackage{amsthm}
\usepackage{dsfont}
\usepackage{bm}

\usepackage[section] {placeins}
\usepackage[section] {placeins}

\usepackage{cases}

\usepackage{algorithm}
\usepackage{algorithmic}

\usepackage{color}
\usepackage{mathrsfs}
\usepackage{amsmath}

\usepackage{stfloats}

\usepackage[dvips]{graphicx}
\usepackage{psfrag}

\usepackage{graphicx}

\usepackage{enumerate}

\usepackage{slashbox}
\usepackage{footnote}

\newtheorem{theorem}{Theorem}
\theoremstyle{definition}
\theoremstyle{remark}

\usepackage{cite}

   \DeclareGraphicsExtensions{.eps,.pdf}

\ifCLASSOPTIONcompsoc
\usepackage[tight,normalsize,sf,SF]{subfigure}
\else
\usepackage[tight,footnotesize]{subfigure}
\fi

\usepackage[english]{babel}

\usepackage{algorithmic}

\begin{document}

\title{D2D-based V2V Communications with Latency and Reliability Constraints$^{*}$}
\author{%
\vspace{-0.5cm}
\IEEEauthorblockA{}{Wanlu Sun, Erik G.~Str\"{o}m, Fredrik Br\"annstr\"om, Yutao Sui, and Kin Cheong Sou}
\IEEEauthorblockA{Department of Signals and Systems, Chalmers University of Technology, Gothenburg, Sweden\\
\emph{\{wanlu,~erik.strom,~fredrik.brannstrom,~sui.yutao,~kincheong.sou\}@chalmers.se}\\\vspace{-1.0cm}
\thanks{$^{*}$This is a revised version of the article found in IEEEXplore. The revisions are related to the description of Modified [8] in Section~\ref{baseline}, and the simulation results of Modified [7] in Fig.~\ref{C_rate_10_5_100_2} and Fig.~\ref{C_rate_10_30_100_3}.}\\
\thanks{This work has been supported in part by the Swedish Research Council project 2011-5824. Part of this work has been performed in the framework of the FP7 project ICT-317669 METIS, which is partly funded by the EU. The authors would like to acknowledge the contributions of their colleagues in METIS, although the views expressed are those of the authors and do not necessarily represent the project.}\\
}
}%

\maketitle
\setlength{\baselineskip}{1.02em}

\begin{abstract}
Direct device-to-device (D2D) communication has been proposed as a possible enabler for vehicle-to-vehicle (V2V) applications, where the incurred intra-cell interference and the stringent latency and reliability requirements are challenging issues. In this paper, we investigate the radio resource management problem for D2D-based V2V communications. Firstly, we analyze and mathematically model the actual requirements for vehicular communications and traditional cellular links. Secondly, we propose a problem formulation to fulfill these requirements, and then a Separate Resource Block allocation and Power control (SRBP) algorithm to solve this problem. Finally, simulations are presented to illustrate the improved performance of the proposed SRBP scheme compared to some other existing methods.
\end{abstract}

\section{Introduction}
\subsection{Motivation}
Recently, vehicle-to-vehicle (V2V) communications have attracted great interest. Usually, these types of applications have a strongly localized nature, i.e., requiring cooperation between vehicles in close proximity. Furthermore, other common features to most applications are real-time requirements, as well as strict requirements on reliability and access availability. For instance, the EU project METIS considers that a maximum end-to-end delay of $5$ ms, with transmission reliability of $99.999\%$ should be guaranteed\cite{METIS_D1.1}.

\par
Current legacy solutions for V2V communications are ad-hoc communications over the 802.11p standard and backend-based communications over the Long Term Evolution (LTE) cellular standard. The main problem with the 802.11p legacy system is that it is mainly optimized for a WLAN-type of environment with no or very low mobility. On the other hand, in LTE systems, as analyzed by \cite{Lottermann_2012}, the performance for vehicular communications is not satisfactory, especially in terms of latency and reliability. Therefore, there is a strong desire of finding better solutions to support V2V communications.

\par

Meanwhile, device-to-device (D2D) communication is identified as one of the technology complements for next generation communication system. In a D2D underlaying cellular infrastructure, two physically close user equipment (UE) devices can directly communicate with each other by sharing the same resources used by regular cellular UEs (C-UEs). Correspondingly, three promising gains, i.e., proximity gain, reuse gain, and hop gain, may be offered \cite{Fodor_2012}.

\par
By comparing the quality of service (QoS) requirements of V2V communications and the potential benefits of D2D communications, it turns out that the direct D2D link can be a possible enabler for V2V application due to the following reasons. Firstly, the localized nature of V2V services is exactly the idea of D2D communications. Moreover, the low latency requirement of V2V applications fits the hop gain of D2D transmissions. Last but not least, V2V's requirement on high reliability is consistent with the proximity gain provided by D2D links. In fact, the D2D underlay network has been proposed as a potential solution for V2V communications in both academic fields \cite{Botsov_2014} and standardization activities \cite{METIS_D1.1}. Nevertheless, using D2D underlay for V2V communications, if performed blindly, may cause significant degradation to system performance due to the interference introduced by resource reuse. Also, to guarantee the required latency and reliability is still a challenge that needs to be tackled for V2V services. Hence, radio resource management (RRM) becomes a key design aspect to enable D2D-based V2V communications.

\subsection{State of the Art}

%
In the context of D2D underlaying systems, one of the most critical challenges is the interference between the primary cellular network and the D2D underlay. To cope with this new interference situation, one crucial issue is the RRM strategy, which includes how the C-UEs and the potential D2D UEs choose the resources to share, and how each UE allocates its transmit power among its used resources. There have been many efforts investigating the RRM problem in such a system. Due to the space limitation, we will only name a few in this field. The interested readers can find more information in excellent survey papers \cite{Fodor_2012,Asadi_2014}, and the references therein.
\par
To maximize the sum rate of the whole network, the authors in \cite{Yu_2011,Zulhasnine_2010,Feng_2013} proposed various algorithms. The work in \cite{Yu_2011} presented mode selection and power control scheme for one D2D link and one C-UE. To generalize the system model, \cite{Zulhasnine_2010} studied the resource allocation for multiple D2D links and C-UEs. Recently, more advanced mathematical techniques have been exploited in RRM problems. In particular, a three-step heuristic resource allocation and power adaptation scheme was derived in \cite{Feng_2013} to maximize the sum rate while guaranteeing the QoS requirements for both D2D users and C-UEs.

\par
In the current research field of RRM for D2D underlaying cellular networks, most existing works aim to maximize the sum rate and prioritize cellular links. Whereas, the D2D underlay is considered as opportunistic when their interference to the cellular links is controlled at acceptable levels. As a result, the schemes proposed for traditional D2D system cannot work for V2V communications (in particular for safety applications) since they usually have strict requirements on latency and reliability but small message payload.

\par
Furthermore, the majority of the literature assume that the eNB is aware of the full instantaneous channel state information (CSI) of all the cellular and D2D links, which is too optimistic for fast moving D2D-based V2V communications, where the vehicle related channels change rapidly.



\subsection{Contributions}
In this work, a Separate Resource Block (RB) allocation and Power control (SRBP) scheme is proposed for the uplink resource sharing in D2D-based V2V communications. From now on, we denote the D2D-based V2V users as vehicular UEs (V-UEs). The main contributions are summarized as follows.
\begin{itemize}
\item We investigate the actual QoS requirements for both C-UEs and V-UEs, and formulate them mathematically.
\item We propose a problem formulation to conduct the RB allocation and power control in D2D-based V2V communications. In the proposed problem, under the condition of satisfying the V-UEs' strict requirements on latency and reliability, we aim at maximizing the C-UEs' sum rate with certain fairness consideration.
\item To solve this problem, we propose a heuristic two-stage SRBP scheme, which is a long-term RRM method, and thus only requires the availability of slow channel fading effects at the eNB.
\end{itemize}


\section{System Model}

\subsection{System Model}
\label{model_interference}
We consider a single cell environment where $M'$ C-UEs and $K'$ V-UEs (counted in terms of the transmitters) share the available uplink radio resources, and the D2D underlay is only used by V-UEs. In general, broadcasting strategies are used for vehicular communication. In this paper, we consider the least favorable receiving vehicle inside the intended broadcast region of each transmitting vehicle, i.e., the vehicle has the smallest channel gain from the transmitting vehicle. The user sets for C-UEs and V-UEs are $\mathcal{M}'\triangleq\{1,2,...,M'\}$ and $\mathcal{K}'\triangleq\{1,2,...,K'\}$, respectively. The whole uplink frequency bandwidth is divided into $F$ subbands with $\mathcal{F}\triangleq \{1,2,...,F\}$ for each scheduling time unit. One subband over one scheduling time unit is defined as one RB. The C-UEs use orthogonal RBs to communicate with the eNB, and the V-UEs use orthogonal RBs among each other. However, an RB can be used by both a C-UE and a V-UE. In this way, interference between the V2V and cellular transmissions will occur.

\par
\begin{figure}[t]
\centering
\psfrag{m}[][][0.7]{$m'$th C-UE}
\psfrag{k}[][][0.7]{$k'$th V-UE Tx}
\psfrag{l}[][][0.7]{$k'$th V-UE Rx}
\psfrag{a}[][][0.7]{$k'$}
\psfrag{N}[][][0.7]{$F$}
\psfrag{a}[][][0.7]{$H'_{m'}$}
\psfrag{d}[][][0.7]{$H_{k'}$}
\psfrag{b}[][][0.7]{$G_{m'k'}$}
\psfrag{c}[][][0.7]{$G'_{k'}$}
\includegraphics[width=2.1in]{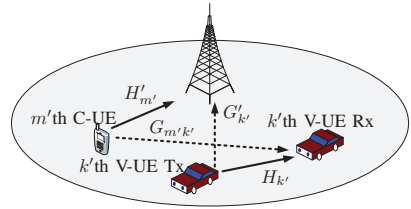}
\caption{Interference between V2V and cellular communications}
\label{system_model}
\vspace{-0.6cm}
\end{figure}

Fig.~\ref{system_model} illustrates the interference situation. Assume the $m'$th C-UE and the $k'$th V-UE are using the same RB. Then they will cause intra-cell interference to each other. $H'_{m'}$ and $H_{k'}$ are the effective channel power gains of the desired transmissions for the $m'$th C-UE and the $k'$th V-UE, respectively. $G_{m'k'}$ denotes the gains of the interference channel from the $m'$th C-UE to the $k'$th V-UE receiver, and $G'_{k'}$ represents the interference channel gain from the $k'$th V-UE to the eNB. To perform RRM, the eNB needs CSI (at least with certain level) for all these involved links, where $H'_{m'}$ and $G'_{k'}$ can be measured at the eNB itself, but $H_{k'}$ and $G_{m'k'}$ have to be measured at the corresponding V-UE receiver and then reported back to the eNB. 

\subsection{Time Scale and Channel Acquisition for RRM}
\label{timescale}
Another potential advantage of D2D communication is to offload the eNB scheduler \cite{Fodor_2012}. To indeed achieve this offloading gain, the time scale of interactions between the eNB and D2D UEs should be much longer than the traditional LTE scheduling time interval ($1$ ms). Furthermore, when D2D communications are used for V2V services, the channels related to V-UEs change very fast. In this case, if the eNB wants meaningful short-term RRM, such as every millisecond, the V-UEs need to report their channels (i.e., $H_{k'}$ and $G_{m'k'}$) every millisecond, which will cause huge overhead. For these two reasons, we claim that the eNB should do long-term, e.g., a few hundred milliseconds, RRM for D2D-based V2V communications. Long-term RRM can also be beneficial for V-UEs that are temporarily out of coverage, as it guarantees resources for these V-UEs.
\par
Regarding channels related to V-UEs, during the considered long-term time period, slow fading effects including path loss and shadowing are quite similar and correlated, but the small scale fading (SSF) changes very fast due to high mobility. Therefore, the available channel information at the eNB should only take the slow fading effects into account since the RRM results must be valid for the next few hundred milliseconds. In this way, the V-UEs merely need to report the slow fading related channel information to the eNB every few hundred milliseconds, which gives an acceptable signaling overhead. 

\section{Requirements on V-UEs and C-UEs}
\label{requirements}
When we are dealing with users having different types of interests, we should consider different requirements for them  \cite{Fodor_2004,Fodor_2001,Fodor_2002}. Undoubtedly, V-UEs and traditional C-UEs ought to have different QoS requirements. In this section, we will clarify our real goal, and mathematically formulate the requirements of V-UEs and C-UEs.

\subsection{Requirements for V-UEs}
\label{requirements_v}
V2V services usually have stringent latency and reliability requirements but are less interested in high data rate. Hence, their requirements can be modeled as strict constraints in our formulation. Now we will study how to consider these requirements mathematically.
\par
{Due to the delay constraints in V2V communications, the RBs assigned to each V-UE should be in a limited time span. Besides, the considered frequency bandwidth is also limited. Hence, the number of RBs that are used for each V-UE's transmission is limited. As analyzed in \cite{Caire_1999}, when assuming a finite number of RBs $E^{\textrm{all}}_{k'}$ for the $k'$th V-UE's transmission, the outage probability evaluated at a required number of bits $N_{k'}$ is defined as
\begin{align}
p^{\textrm{out}}_{k'}\triangleq\textrm{Pr}\left\{\sum_{i=1}^{E^{\textrm{all}}_{k'}}\rho \log_2\left(1+\gamma_i\right)< N_{k'}\right\},\label{op1_def}
\end{align}
where $\gamma_i\triangleq{\bar{P}^{\textrm{r}}_{i}}|h_i|^2/(\sigma^2+\bar{S}^{\textrm{r}}_{i}|g_i|^2)$ is the instantaneous Signal to Interference plus Noise Ratio (SINR) on the $i$th RB; $\bar{P}^{\textrm{r}}_{i}$ and $\bar{S}^{\textrm{r}}_{i}$ are average received power from the desired and interfering users, respectively; $h_i$ and $g_i$ are random variables which represent the SSF effects of the corresponding desired channel and interference channel; $\sigma^2$ is the noise power; and $\rho$ is the number of complex symbols per RB. Then, similar to \cite{METIS_D1.1}, the reliability requirement is interpreted from the perspective of outage probability and can be expressed as
\begin{align}
p^{\textrm{out}}_{k'}\leq p_\textrm{o},\label{op1}
\end{align}
where $p_\textrm{o}$ is the maximum tolerable outage probability.
}
\par
{Furthermore, as explained in Section~\ref{timescale}, the eNB only requires and possesses the slow fading effects of channels. In this case, the reliability constraint considered by the eNB for implementing RRM should only involve the slow fading information. To do so, we will replace the requirement in (\ref{op1}) by a more restrict requirement. We first upper-bound $p^{\textrm{out}}_{k'}$ by replacing $\gamma_i$ with $\bar{\gamma}_i|h_i|^2/(1+|g_i|^2)$, where $\bar{\gamma}_i\triangleq{\bar{P}^{\textrm{r}}_{i}}/(\sigma^2+\bar{S}^{\textrm{r}}_{i})$ only including the slow fading knowledge. In this way, if the upper-bounded probability is smaller than $p_\textrm{o}$, the original inequality in (\ref{op1}) is always satisfied. Then, we further restrict the new outage probability requirement into the following two constraints,
\begin{align}
&\textrm{Pr}\left\{\sum_{i=1}^{E^{\textrm{all}}_{k'}}\rho \log_2\left(1+\bar{\gamma}_{k'}^{\textrm{T}}\frac{|h_i|^2}{1+|g_i|^2}\right)< N_{k'}\right\}\leq p_\textrm{o}\label{op2_1}\\
&\bar{\gamma}_i\geq \bar{\gamma}_{k'}^{\textrm{T}}, \quad \forall i=1,2,...,E^{\textrm{all}}_{k'}. \label{op2_2}
\end{align}
Constraints (\ref{op2_1}) and (\ref{op2_2}) mean that, for the $k'$th V-UE, by deriving $\bar{\gamma}_{k'}^{\textrm{T}}$ from (\ref{op2_1}) and forcing the actual $\bar{\gamma}_i$ on each used RB larger than $\bar{\gamma}_{k'}^{\textrm{T}}$, we can guarantee that (\ref{op1}) will be satisfied. Note that $\bar{\gamma}_i$ contains our decision variables that will be introduced later. From now on, with a slight abuse of terminology, we denote (\ref{op2_2}) as the SINR constraint.
}

\par
Now, for a given $\rho$, $N_{k'}$, $p_\textrm{o}$, and the probability density function (pdf) of $h_i$ as well as $g_i$, we can derive $\bar{\gamma}_{k'}^{\textrm{T}}$ from $E^{\textrm{all}}_{k'}$, e.g., by Monte Carlo (MC) simulation methods. The choice of $E^{\textrm{all}}_{k'}$ depends on the traffic load of the network, and its joint optimization with other parameters in problem (\ref{1TTI_sim_2}) is left for future work.

\par
Moreover, to meet the latency constraint, the $E^{\textrm{all}}_{k'}$ RBs have to be allocated within the RB region $F\times L_{\textrm{tol}}$, where $L_{\textrm{tol}}$ is the maximum tolerable latency of V2V communications in terms of the number of scheduling time units. Notice that in reality we have multiple V-UEs which may appear at different time. So it is hard to find a common two dimensional region to implement RB allocation for all the V-UEs. Therefore, we will reduce the two dimensional RB allocation problem into a sequence of one-dimensional problems, i.e., only over frequency. Correspondingly, the requirements on latency and reliability become
\begin{align}
&E_{k'}=\lceil E^{\textrm{all}}_{k'}/L_{\textrm{tol}}\rceil \label{op3_1}\\
&\bar{\gamma}_i\geq \bar{\gamma}_{k'}^{\textrm{T}}, \quad \forall i=1,2,...,E_{k'}, \label{op3_2}
\end{align}
where $E_{k'}$ is the number of RBs allocated to the $k'$th V-UE during each scheduling time unit, and we have $\sum_{k'=1}^{K'}E_{k'}\leq F$. The calculation of $E_{k'}$ in (\ref{op3_1}) ensures that at least $E^{\textrm{all}}_{k'}$ RBs will be allocated to the $k'$th V-UE within $L_{\textrm{tol}}$ time units.
\par
In this way, we transformed the original V2V requirements on latency and reliability into the constraints on $E_{k'}$ and $\bar{\gamma}_{k'}^{\textrm{T}}$. {To summarize, if the $k'$th V-UE is assigned $E_{k'}$ RBs during each time unit where the actual $\bar{\gamma}_i$ on the $i$th used RB is larger than $\bar{\gamma}_{k'}^{\textrm{T}}$, then the original latency and reliability requirements can be satisfied for this V-UE.}

\subsection{Requirements for C-UEs}
In contrast to V2V safety communications, for most type of the cellular traffic, the latency requirement is less strict, and the system usually aims at maximizing the sum throughput under certain fairness considerations. Therefore, the maximization of the C-UEs' sum rate (as defined in Section~\ref{problem_formulation}) can be formulated as the objective in our problem.
\par
With regard to fairness, here we assume the proportional bandwidth fairness \cite{Sadr_2007} among C-UEs that means the number of RBs allocated to the $m'$th C-UE $E'_{m'}$ during one scheduling time unit is given for all $m'\in\mathcal{M}'$ and $\sum_{m'=1}^{M'}E'_{m'}=F$.

\section{Problem Formulation}
\label{problem_formulation}
In this section, we detail the RRM problem formulation for D2D-based V2V communications, which fulfills the requirements of V-UEs and C-UEs at the same time. To summarize, our objective is to maximize the C-UEs' sum rate with fairness considerations, under the condition of satisfying V-UEs' requirements on latency and reliability, i.e., constraints (\ref{op3_1}) and (\ref{op3_2}).

\par
For notational convenience in Section~\ref{proposed_SRBP}, we introduce the concepts of sub-users and extended user sets. Firstly, we include one dummy V-UE, i.e., the $(K'+1)$th V-UE, with the number of allocated RBs being $E_{K'+1}=F-\sum_{k'=1}^{K'}E_{k'}$. Besides, to complete the dummy V-UE related information, we let $H_{K'+1}= +\infty$, $G'_{K'+1}=0$, $\bar{\gamma}_{K'+1}^{\textrm{T}}= 0$, and $G_{m'(K'+1)}=0$ for all $m'\in \mathcal{M}'$. Then, we divide the $k'$th V-UE into $E_{k'}$ sub-V-UEs for all $k'\in \tilde{\mathcal{K}}\triangleq\{\mathcal{K}'\cup \{K'+1\}\}$, and divide the $m'$th C-UE into $E'_{m'}$ sub-C-UEs for all $m'\in \mathcal{M}'$, where each sub-user uses exactly one RB. Moreover, we define two extended user sets $\mathcal{K}\triangleq\{1,2,...,F\}$ and $\mathcal{M}\triangleq\{1,2,...,F\}$ for sub-V-UEs and sub-C-UEs, respectively. In this way, we have $K=M=F$, where $K=|\mathcal{K}|$ and $M=|\mathcal{M}|$. Here $|\mathcal{X}|$ denotes the cardinality of the set $\mathcal{X}$. To relate the original user sets and the extended user sets, we define $\hat{k}$: $\mathcal{K}\rightarrow \tilde{\mathcal{K}}$ such that $k'=\hat{k}(k)$ is the V-UE to which the sub-V-UE $k$ belongs. Similarly, the function $\hat{m}$: $\mathcal{M}\rightarrow\mathcal{M}'$ is such that $m'=\hat{m}(m)$ is the C-UE to which the sub-C-UE $m$ belongs.

\par

Based on the above definitions, the problem is mathematically formulated as maximizing the C-UEs' sum rate, i.e.,
\begin{align}
&\max \sum_{m=1}^{M}\sum_{f=1}^F \log_2\left(1+\frac{{S_{fm}}H^{'}_{\hat{m}(m)}}{\sigma^2+\sum_{k=1}^{K}{P_{fk}}G^{'}_{\hat{k}(k)}}\right)\label{1TTI_sim_2}
\end{align}
subject to
\begin{align}
&{q_{fk}}\in\{0,1\}, \quad {l_{fm}}\in\{0,1\},\quad \forall f,k,m\tag{\theequation a}\label{c2_1}\\
&\sum_{f=1}^F\sum_{k,\hat{k}(k)=k'} {P_{fk}}\leq P_{\textrm{max}}^{\textrm{V}},\quad\forall k'\quad \tag{\theequation b}\label{c2_8}\\
&\sum_{f=1}^F\sum_{m,\hat{m}(m)=m'} {S_{fm}}\leq P_{\textrm{max}}^{\textrm{C}},\quad\forall m'\quad \tag{\theequation c}\label{c2_7}\\
&0\leq {P_{fk}}\leq P_{\textrm{max}}^{\textrm{V}}{q_{fk}}, \quad \forall f,k\quad \tag{\theequation d}\label{c2_3}\\
&0\leq {S_{fm}}\leq P_{\textrm{max}}^{\textrm{C}}{l_{fm}}, \quad \forall f,m \quad \tag{\theequation e}\label{c2_4}\\
&\sum_{k=1}^{K} {q_{fk}}= 1, \quad \sum_{m=1}^{M}{l_{fm}}= 1,\quad \forall f\quad \tag{\theequation f}\label{c2_5}\\
&\sum_{f=1}^F q_{fk}= 1,\quad\sum_{f=1}^F l_{fm}= 1, \quad \forall k,m\quad \tag{\theequation g}\label{c2_10}\\
&\frac{{P_{fk}}H_{\hat{k}(k)}}{\sigma^2+\sum\nolimits_{m=1}^{M}{S_{fm}}G_{\hat{m}(m)\hat{k}(k)}}\geq{q_{fk}} \bar{\gamma}_{\hat{k}(k)}^{\textrm{T}}, \quad \forall f,k\tag{\theequation h}\label{c2_11}
\end{align}
where $f\in \mathcal{F}$, $k\in \mathcal{K}$, $m\in \mathcal{M}$, $k'\in \tilde{\mathcal{K}}$, $m'\in \mathcal{M}'$, $q_{fk}$ ($l_{fm}$) is a binary variable equal to $1$ if the $k$th sub-V-UE ($m$th sub-C-UE) is assigned to the $f$th RB and $0$ otherwise; $P_{fk}$ ($S_{fm}$) is the transmit power of the $k$th sub-V-UE ($m$th sub-C-UE) on the $f$th RB. (\ref{c2_8}) and (\ref{c2_7}) represent the max transmit power constraints for each V-UE and C-UE, respectively. Constraint (\ref{c2_3}) (constraint (\ref{c2_4})) forces the transmit power of the $k$th sub-V-UE (the $m$th sub-C-UE) on the $f$th RB to be $0$ in case $q_{fk}=0$ ($l_{fm}=0$). (\ref{c2_5}) guarantees the orthogonal RB allocation among V-UEs and among C-UEs. (\ref{c2_10}) ensures the number of RB assigned to each sub-V-UE and each sub-C-UE is exactly one. Last but not least, (\ref{c2_11}) enforces the SINR constraint for each sub-V-UE, where the LHS is interpreted as $\bar{\gamma}_{k}$.
\par
In problem (\ref{1TTI_sim_2}), the inputs are $F$, $M'$, $K'$, $E'_{m'}$, $E_{k'}$, $\sigma^2$, $\bar{\gamma}_{k'}^{\textrm{T}}$, $P_{\textrm{max}}^{\textrm{V}}$, $P_{\textrm{max}}^{\textrm{C}}$, $H^{'}_{m'}$, $G^{'}_{k'}$, $H_{k'}$, and $G_{m'k'}$; and the outputs (also the optimization variables) are $q_{fk}$, $l_{fm}$, $P_{fk}$, and $S_{fm}$ for all $f\in \mathcal{F}$, $k\in \mathcal{K}$, and $m\in \mathcal{M}$.

\par
Unfortunately, the proposed problem formulation in (\ref{1TTI_sim_2}) is a mixed-integer non-linear program which is computationally intractable. Therefore, heuristic solutions will be applied here. 

\section{The Proposed SRBP Algorithm}
\label{proposed_SRBP}
In this section, we will propose an SRBP scheme to solve problem (\ref{1TTI_sim_2}). There are two stages in the SRBP algorithm. Firstly, assuming equal power allocation, the eNB allocates RBs to both V-UEs and C-UEs in an optimal and time efficient way by transforming the RB allocation problem into an maximum weight matching (MWM) problem for bipartite graphs. See \cite{Goemans_2009} for general background on MWM. Secondly, based on the RB allocation results from the first stage, the eNB further optimally adjusts the transmit power for each V-UE and C-UE. This is realized via transforming the power control problem into a convex optimization problem, which can be solved by, e.g., an interior point method. In this way, even though the proposed SRBP method is heuristic by dividing the whole process into two stages, we can achieve the optimal solution in both stages, which to some extent promises good performance of the SRBP algorithm, which is confirmed by numerical results in Section~\ref{simulation_results}.

\subsection{RB Allocation}
\label{RB_allocation}
Initially, we assume equal power allocation for each V-UE and C-UE on each of their used RBs, i.e., for the $k'$th V-UE, the power on each of its used RBs is $P_{k'}^\textrm{V}\triangleq P_{\textrm{max}}^{\textrm{V}}/E_{k'}$. Likewise, for the $m'$th C-UE, the power on each of its used RBs is $P_{m'}^\textrm{C}\triangleq P_{\textrm{max}}^{\textrm{C}}/E'_{m'}$. In this way, problem (\ref{1TTI_sim_2}) reduces to an integer program that is denoted as the RB allocation problem, where the optimization variables are $q_{fk}$ and $l_{fm}$ for all $f\in \mathcal{F}$, $k\in \mathcal{K}$, and $m\in \mathcal{M}$. The RB allocation problem is similar to the formulation in (\ref{1TTI_sim_2}) by replacing $P_{fk}$ and $S_{fm}$ with $q_{fk}P_{\hat{k}(k)}^\textrm{V}$ and $l_{fm}P_{\hat{m}(m)}^\textrm{C}$ respectively, as well as eliminating constraints (\ref{c2_8}), (\ref{c2_7}), (\ref{c2_3}), and (\ref{c2_4}).

\par
In the following, we will propose Theorem \ref{RB} to transform the RB allocation problem into an MWM problem for bipartite graphs, which can be then solved optimally by the Hungarian algorithm \cite{Goemans_2009}.

\begin{theorem}
\label{RB}
The RB allocation problem can be transformed into the following equivalent optimization problem.
\begin{align}
&\max \sum_{m=1}^{M}\sum_{k=1}^{K} x_{mk}\left( \log_2\left(1+\frac{P_{\hat{m}(m)}^\textrm{C}H^{'}_{\hat{m}(m)}}{\sigma^2+P_{\hat{k}(k)}^\textrm{V}G^{'}_{\hat{k}(k)}}\right)\right.\label{RB_problem_MWM}\\
&\left.+\phi\min\left\{\frac{P_{\hat{k}(k)}^\textrm{V}H_{\hat{k}(k)}}{\sigma^2+P_{\hat{m}(m)}^\textrm{C}G_{\hat{m}(m)\hat{k}(k)}}-\bar{\gamma}_{\hat{k}(k)}^{\textrm{T}},0\right\}\right)\notag
\end{align}
$\textnormal{subject to}$
\begin{align}
& x_{mk}\in\{0,1\}, \quad \forall m\in \mathcal{M},k\in \mathcal{K}\tag{\theequation a}\label{RB_problem_bi_a}\\
&\sum_{k=1}^{K} x_{mk}=1, ~\forall m\in \mathcal{M}, \quad \sum_{m=1}^{M} x_{mk}=1, ~\forall k\in \mathcal{K} \tag{\theequation b}\label{RB_problem_bi_b}
\end{align}
where $\phi$ is the penalty coefficient, which should be a large enough value to force the SINR constraints in (\ref{c2_11}) to be satisfied under the assumption of equal power allocation, if possible.
\end{theorem}
\begin{IEEEproof}
Due to the space limitation, a rigorous proof is not given here and will be reported in our future work. The key idea is to firstly define $x_{mk}\triangleq\sum_{f=1}^{F} q_{fk}l_{fm}$; secondly involve the SINR constraints in (\ref{c2_11}) as penalties into the objective function; and finally show the equivalence between the two objective functions as well as the two sets of constraints, respectively.
\end{IEEEproof}

\par
In fact, problem (\ref{RB_problem_MWM}) has its own meaning. Based on the definition of sub-users, the binary variable $x_{mk}$ is equal to $1$ if the $m$th sub-C-UE and $k$th sub-V-UE are sharing the same RB and is equal to $0$ otherwise. Also, each sub-C-UE is required to share the same RB with exactly one sub-V-UE, and vice versa. Then, problem (\ref{RB_problem_MWM}) is maximizing the sum rate of sub-C-UEs under the condition of satisfying the SINR constraint for each sub-V-UE. Furthermore, as analyzed in Section~\ref{timescale}, the available channel information of all the involved links is the same in the whole considered frequency range. Hence, after pairing a sub-C-UEs with the corresponding sub-V-UEs, there is no difference which RB each pair is using as long as different pairs are using orthogonal RBs.


\par
Problem (\ref{RB_problem_MWM}) fits perfectly into the MWM problem for bipartite graphs. Thus, the Hungarian algorithm \cite{Goemans_2009} is an efficient way to solve problem (\ref{RB_problem_MWM}) within polynomial time, where the number of operations is upper bounded by $O(F^3)$ \cite{Goemans_2009}.

\subsection{Power Control}
\label{power_control}
The second stage of the proposed SRBP algorithm is power control. According to $x^*_{mk}$ obtained by solving problem (\ref{RB_problem_MWM}), the power control problem can be formulated as
\begin{align}
&\max \sum_{m=1}^{M} \sum_{k=1}^{K}x^*_{mk}\log_2\left(1+\frac{{S_{m}}H^{'}_{\hat{m}(m)}}{\sigma^2+{P_{k}}G^{'}_{\hat{k}(k)}}\right)\label{PC}
\end{align}
subject to
\begin{align}
&{S_{m}}\geq 0, \quad {P_{k}}\geq 0,\quad \forall m\in \mathcal{M}, \in \mathcal{K} \tag{\theequation b}\label{PC_posi}\\
& \sum_{m,\hat{m}(m)=m'} {S_{m}}\leq P_{\textrm{max}}^{\textrm{C}},\quad\forall m'\in \mathcal{M}' \tag{\theequation c}\label{PC_PC}\\
& \sum_{k,\hat{k}(k)=k'} {P_{k}}\leq P_{\textrm{max}}^{\textrm{V}},\quad\forall k'\in \mathcal{K}' \tag{\theequation d}\label{PC_PV}\\
& \frac{{P_{k}}H_{\hat{k}(k)}}{\sigma^2+\sum_{m=1}^{M}x^*_{mk}S_{m}G_{\hat{m}(m)\hat{k}(k)}}\geq \bar{\gamma}_{\hat{k}(k)}^{\textrm{T}}, \quad \forall k\in \mathcal{K}\tag{\theequation e}\label{PC_SINR}
\end{align}
where the optimization variables are $\{{S_{m}}\}_{m=1}^{M}$ and $\{{P_{k}}\}_{k=1}^{K}$. Since the objective function in (\ref{PC}) is not concave with respect to  $\{{P_{k}}\}_{k=1}^{K}$, this problem is not convex. Nevertheless, notice that the objective function is monotonically nonincreasing in terms of $\{{P_{k}}\}_{k=1}^{K}$, and thus we can eliminate $\{{P_{k}}\}_{k=1}^{K}$ by achieving the equalities of the constraints (\ref{PC_SINR}). Then, the remaining problem is transformed into an equivalent convex optimization problem which can be solved optimally by an interior point method.

\section{Performance Evaluation}
\subsection{Scenarios and Parameters}
We assume a single cell outdoor system with a carrier frequency of $800$ MHz and that each RB has a bandwidth of $180$ KHz for the uplink communication. In particular, we consider test case (TC) $2$ \cite{METIS_D6.1} defined by METIS, which describes an urban environmental model similar with the Manhattan grid layout. In this topology, the entire region is a $444$ m $\times$ $444$ m square and the size of each building is $120$ m $\times$ $120$ m.

\par
The used channel models are specified by \cite{METIS_D6.1}, which describes the large scale modeling for different propagations scenarios (PSs). Specifically, we refer to PS$\#3$ in \cite{METIS_D6.1} for the links connected to the eNB (i.e., $H'_{m'}$ and $G'_{k'}$); and PS$\#9$ in \cite{METIS_D6.1} for the links between UEs (i.e., $H_{k'}$ and $G_{m'k'}$).

\par
Simulation parameters are summarized as follows. $\rho=84$, $P_{\textrm{max}}^{\textrm{V}}=P_{\textrm{max}}^{\textrm{C}}=24$ dBm. Besides, the antenna height is $26$ m at the eNB and is $1.5$ m at each UE. The distance between two communicating V-UEs we consider here is $18$ m. Also, the noise floor is $-117$ dBm at the eNB and each V-UE. The SSF of the channels is assumed to be Rayleigh distribution with unit power gain. Finally, one scheduling time unit (i.e., the time period of one RB) is $0.5$ ms and the time scale of RRM is $100$ ms.
\subsection{Performance Metric and Compared Schemes}
\label{baseline}
We base our evaluation on three metrics:
\begin{itemize}
\item C-UEs' sum rate when SSF is disregarded;
\item cumulative distribution function (CDF) of C-UEs' sum rate;
\item CDF of one V-UE's transmitted bits within $5$ ms, i.e., the left hand side of the inner inequality in (\ref{op1_def}).
\end{itemize}
The last two metrics are evaluated when considering SSF in simulations.
\par
Moreover, to let the comparison be as fair as possible, we will make modifications to existing baseline methods as follows.
\par
1) Modified \cite{Zulhasnine_2010}. In \cite{Zulhasnine_2010}, the eNB selects the C-UE with highest desired channel gain to share its RB with the V-UE which suffers the lowest interference from this C-UE. The method is executed with the max power. To make the scheme fit our framework where each UE can use multiple RBs, we replace the concepts of C-UE and V-UE with the concepts of sub-C-UE and sub-V-UE. Correspondingly, the max power constraints become $P_{m'}^\textrm{C}$ and $P_{k'}^\textrm{V}$ for each sub-C-UE and each sub-V-UE. Furthermore, to meet the SINR constraint for each sub-V-UE, we simply decrease the transmit power of the corresponding sub-C-UEs until the SINR constraint is just satisfied.
\par
2) Modified \cite{Feng_2013}. In \cite{Feng_2013}, a three-step scheme is derived to maximize the sum rate of both C-UEs and V-UEs. Here we have two modifications. Firstly, like in Modified \cite{Zulhasnine_2010}, we use the concepts of sub-C-UE and sub-V-UE. Besides, we change the objective in the second step of the algorithm from maximizing the sum rate of both C-UEs and V-UEs into maximizing the sum rate of C-UEs.
\par
3) Optimal solution to problem (\ref{1TTI_sim_2}), which is achieved by firstly conducting the exhaustive search over all the RB allocation possibilities, and then implementing the optimal power control for each RB allocation result. Due to its exponentially increased complexity, we only simulate the optimal solution for $F=4$.

\subsection{Simulation Results}
\label{simulation_results}
Based on the requirements given by METIS \cite{METIS_D1.1}, we have $N_{k'}=12800$ bits, $p_\textrm{o}=10^{-5}$ (i.e., a transmission reliability of $99.999\%$), and $L_{\textrm{tol}}=10$ (i.e., a latency requirement of $5$ ms). As analyzed in Section~\ref{requirements_v}, the relationship between $E^{\textrm{all}}_{k'}$ and $\bar{\gamma}_{k'}^{\textrm{T}}$ can be derived from (\ref{op2_1}) through a MC method. Then $E_{k'}$ can be calculated via (\ref{op3_1}). In this way, some possible values of $\{E_{k'},\bar{\gamma}_{k'}^{\textrm{T}}\textrm{[dB]}\}$ are $\{2,34.3\}$, $\{3,24.9\}$, and $\{4,19.82\}$.

\par
\begin{figure}[t]
\centering
\psfrag{f}[][][0.85]{CDF}
\psfrag{c}[][][0.85]{bit/s/Hz}
\psfrag{Proposed SRBP pppp}[][][0.85]{Proposed SRBP, $6.49$}
\psfrag{Modified 10 ooooo}[][][0.85]{Modified \cite{Zulhasnine_2010}, $1.70$}
\psfrag{Modified 12 eeeee}[][][0.85]{Modified \cite{Feng_2013}, $5.64$}
\psfrag{optimal ooooo}[][][0.85]{Optimal, $6.56$}
\includegraphics[width=3.4in]{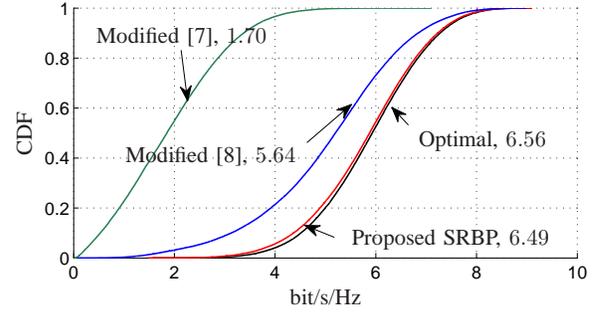}
\caption{Sum rate of C-UEs. $F=4$, $M'=4$, $K'=2$, $E'_{m'}=1$, and $E_{k'}=2$.}
\label{C_rate_4_2_4_2}
\vspace{-0.4cm}
\end{figure}

Fig.~\ref{C_rate_4_2_4_2} compares C-UEs' sum rates of different schemes when $F=4$, which is plotted to show the performance gap with the optimal solution. The numbers in the legend represent the achieved rates when the SSF is not taken into account. In other words, the rate when the utilized channel knowledge in the four RRM methods matches the actual channel in the simulations. Besides, the CDF curves show C-UEs' sum rates when the SSF is also involved in simulated channels. It can be seen that these long-term RRM schemes do not incur big difference on the average performance when being applied to realistic channels with SSF effects. Regarding the evaluation of different methods, the performance degradation of the proposed SRBP is fairly slight compared to the optimal solution. This is because the SRBP can actually lead to the optimal result in each step, even though it is a heuristic two-step scheme. On the other hand, Modified \cite{Zulhasnine_2010} and Modified \cite{Feng_2013} exhibit worse sum rates.


\begin{figure}[t]
\centering
\psfrag{f}[][][0.85]{CDF}
\psfrag{c}[][][0.85]{bit/s/Hz}
\psfrag{S}[][][0.85]{Proposed SRBP, $3.98$}
\psfrag{Z}[][][0.85]{Modified \cite{Zulhasnine_2010}, $2.09$}
\psfrag{D}[][][0.85]{Modified \cite{Feng_2013}, $3.71$}
\includegraphics[width=3.4in]{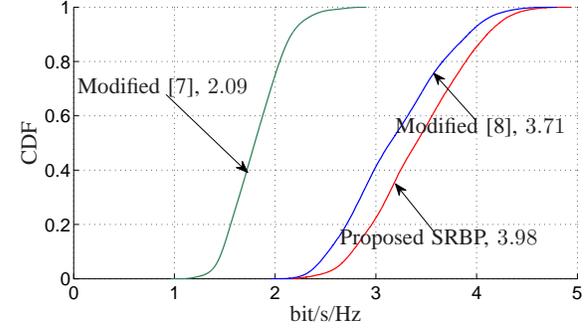}
\caption{Sum rate of C-UEs. $F=100$, $M'=10$, $K'=5$, $E'_{m'}=10$, and $E_{k'}=2$.}
\label{C_rate_10_5_100_2}
\vspace{-0.4cm}
\end{figure}

\begin{figure}[t]
\centering
\psfrag{f}[][][0.85]{CDF}
\psfrag{c}[][][0.85]{bits}
\psfrag{Proposed SRBP}[][][0.85]{Proposed SRBP}
\psfrag{Modified 10 o}[][][0.85]{Modified \cite{Zulhasnine_2010}}
\psfrag{Modified 12 e}[][][0.85]{Modified \cite{Feng_2013}}
\psfrag{N}[][][0.85]{$N_{k'}$}
\psfrag{p}[][][0.85]{$p_\textrm{o}$}
\includegraphics[width=3.4in]{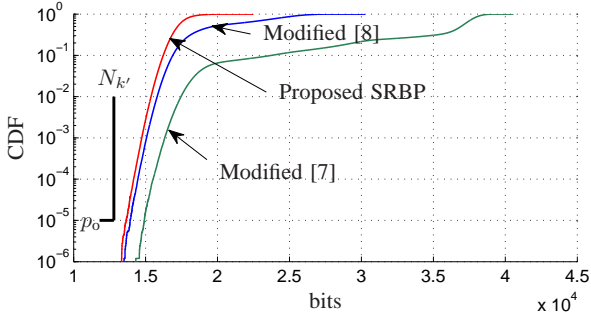}
\caption{Transmitted bits within $5$ ms for each V-UE. $F=100$, $M'=10$, $K'=5$, $E'_{m'}=10$, and $E_{k'}=2$.}
\label{D_rate_10_5_100_2}
\vspace{-0.4cm}
\end{figure}

\par
For a low load scenario, i.e., $K'=5$, Fig.~\ref{C_rate_10_5_100_2} and Fig.~\ref{D_rate_10_5_100_2} illustrate the performances of C-UEs and V-UEs respectively.
\par
In Fig.~\ref{C_rate_10_5_100_2}, C-UEs' sum rates are evaluated. Compared to the SRBP and Modified \cite{Feng_2013}, Modified \cite{Zulhasnine_2010} has obviously degraded performance, which is mainly due to two reasons. Firstly, the Modified \cite{Zulhasnine_2010} prioritize the C-UEs' QoS requirements and aims at maximizing the sum rate of both C-UEs and V-UEs. Thus, there is no QoS guarantee on the V-UEs. In this way, when we conduct the modifications described above to satisfy V-UEs' SINR constraints, the transmit power of C-UEs will be sacrificed and, hence, their rates. Secondly, the scheme (proposed in \cite{Zulhasnine_2010}) itself is a greedy method and can be improved by more advanced techniques. On the other hand, although a sophisticated power adaptation and RB assignment algorithm is utilized in the Modified \cite{Feng_2013}, the proposed SRBP still reveals superiority. This is because the power adjustment among different RBs used by one UE is not considered in \cite{Feng_2013}. However, in the SRBP algorithm, we solve the power control problem optimally under a given RB allocation.
\par
Fig.~\ref{D_rate_10_5_100_2} shows the CDF of the transmitted bits within $5$ ms for one V-UE. It can be seen that the outage probability constraint which represents the QoS requirements on V-UEs is fulfilled for all the three schemes. We stress the fact that there is no need to exceed the requirements for V-UEs. Indeed, the fact that Modified \cite{Zulhasnine_2010} and Modified \cite{Feng_2013} do this to a higher degree than SRBP explains why their C-UE sum rates are worse than SRBP (see Fig.~\ref{C_rate_10_5_100_2}).



\begin{figure}[t]
\centering
\psfrag{f}[][][0.85]{CDF}
\psfrag{c}[][][0.85]{bit/s/Hz}
\psfrag{Proposed SRBP pp}[][][0.85]{Proposed SRBP, $3.94$}
\psfrag{Modified 12 eee}[][][0.85]{Modified \cite{Zulhasnine_2010}, $1.02$}
\psfrag{Modified 10 ooo}[][][0.85]{Modified \cite{Feng_2013}, $2.27$}
\includegraphics[width=3.4in]{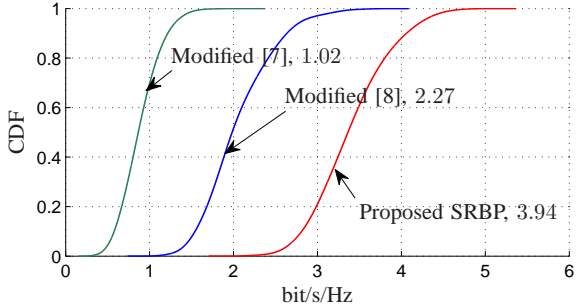}
\caption{Sum rate of C-UEs. $F=100$, $M'=10$, $K'=30$, $E'_{m'}=10$, and $E_{k'}=3$.}
\label{C_rate_10_30_100_3}
\end{figure}

Next, C-UEs' sum rates are compared in Fig.~\ref{C_rate_10_30_100_3} for relatively high load scenario, i.e., $K'=30$. In this case, the performance gap between the Modified \cite{Feng_2013} and the proposed SRBP is more significant, which is attributed to the following reason. The more V-UEs there exist, the more interference the overall C-UEs suffer from. Then, the power control, more specifically, the smart power allocation for one UE on its used RBs, plays a more important role in order to control the interference and improve the entire system performance. Besides, by comparing the performance of the proposed SRBP algorithm in Fig.~\ref{C_rate_10_30_100_3} and Fig.~\ref{C_rate_10_5_100_2}, it is revealed that the SRBP scheme is quite robust to different traffic loads, which further demonstrates its advantage in practice.

\par
Due to the space limitation, here we do not present results to discuss the impact of $E_{k'}$ on the performance. However, our experience is that Modified \cite{Zulhasnine_2010} is more sensitive to changes in $E_{k'}$ than Modified \cite{Feng_2013}. SRBP, on the other hand, is relative insensitive to $E_{k'}$ changes. A more detailed analysis will be given in our future work.


\vspace{-0.2cm}
\section{Conclusion}
Due to the similarity between the QoS requirements of V2V services and the benefits of D2D communications, the direct D2D link can be envisioned as a possible enabler for V2V applications as long as the RRM is designed in a smart way. In this paper, the SRBP scheme is proposed for D2D-based V2V communications, which aims to maximize C-UEs' sum rate under the condition of fulfilling V-UEs' requirements on latency and reliability. Besides, simulation results using realistic channel models are presented to demonstrate the superiority of the proposed SRBP algorithm over some existing schemes.

\bibliographystyle{IEEEtran}
\bibliography{ref_wanlu_V2V}

\end{document}